\documentclass{aa}

\usepackage{txfonts}

\usepackage{graphicx}
\usepackage{color}
\usepackage{url}

\usepackage[]{natbib}
\bibpunct{(}{)}{;}{a}{}{,} % to follow the A&A style

\newcommand{\myfigure}[3]{
  \begin{figure}
    \resizebox{\hsize}{!}{\includegraphics{#1}}
    \caption{#2}
    \label{#3}
  \end{figure}
}

\newcommand{\mytabletwo}[5]{
  \begin{table*}
    \caption{#4}              % title of Table
    \label{#5}      % is used to refer this table in the text
    \centering                                      % used for centering table
    \begin{tabular}{#1}          % centered columns (4 columns)
      \hline\hline                        % inserts double horizontal lines
      #2 \\    % table heading
      \hline                                   % inserts single horizontal line
      #3 % inserting body of the table
      \hline                                             %inserts single line
    \end{tabular}
  \end{table*}
}

\newcommand{\mytable}[5]{
  \begin{table}
    \caption{#4}              % title of Table
    \label{#5}      % is used to refer this table in the text
    \centering                                      % used for centering table
    \begin{tabular}{#1}          % centered columns (4 columns)
      \hline\hline                        % inserts double horizontal lines
      #2 \\    % table heading
      \hline                                   % inserts single horizontal line
      #3 % inserting body of the table
      \hline                                             %inserts single line
    \end{tabular}
  \end{table}
}
\begin{document}
   \title{OmegaWINGS: OmegaCAM@VST observations of WINGS galaxy clusters\thanks{Based on observations made with VST at ESO Paranal Observatory under program ID 88.A-4005, 089.A-0023, 090.A-0074, 091.A-0059, and 093.A-0041.}}
%   \subtitle{Observing strategies and data reduction}
   \author{
     M. Gullieuszik  \inst{1}\and
     B. Poggianti    \inst{1}\and
     G. Fasano       \inst{1}\and
     S. Zaggia       \inst{1}\and
     A. Paccagnella  \inst{1,2}\and
     A. Moretti      \inst{1,2}\and
     D. Bettoni      \inst{1}\and
     M. D'Onofrio    \inst{1}\and
     W. J. Couch     \inst{3}\and
     B. Vulcani      \inst{4}\and
     J. Fritz        \inst{5,6}\and
     A. Omizzolo     \inst{1,7}\and
     A. Baruffolo\inst{1}\and
     P. Schipani\inst{8}\and
     M. Capaccioli\inst{9}\and
     J. Varela \inst{10}
   }
   \institute{
     INAF - Osservatorio astronomico di Padova, Vicolo
     dell'Osservatorio 5, 35122 Padova, Italy
     \and
     Dipartimento di Fisica e Astronomia, Universit\`a degli Studi di
     Padova, Vicolo dell'Osservatorio 3, 35122 Padova, Italy
     \and
     Australian Astronomical Observatory, PO Box 915, North Ryde, NSW
     1670 Australia
     \and
     Kavli Institute for the Physics and Mathematics of the Universe
     (WPI), Todai Institutes for Advanced Study, the University of
     Tokyo, Kashiwa, 277-8582, Japan
     \and
     Sterrenkundig Observatorium Vakgroep Fysica en Sterrenkunde
     Universiteit Gent, Krijgslaan 281, S9 9000 Gent, Belgium
     \and
     Centro de Radioastronom'a y Astrof'sica, UNAM, Campus Morelia, A.P. 3-72, C.P. 58089, Mexico
     \and
     Specola Vaticana, 00120, Vatican City State 
     \and
     INAF - Osservatorio astronomico di Capodimonte, Salita
     Moiariello 16, 80131 Napoli, Italy
     \and
     Dipartimento di Fisica, Universit\`a Federico II, via Cinthia, I-80126 Napoli, Italy
     \and
     Centro de Estudios de Fisica del Cosmos de Aragon, Plaza San Juan,
     1, 44001 Teruel, Spain
   }

\date{Received xxx; accepted xxx}

  \abstract
  {Wide-field observations targeting galaxy clusters at low redshift are complementary to field surveys and provide the local benchmark for detailed studies of the most massive haloes in the local Universe. The Wide-field Nearby Galaxy-cluster Survey (WINGS) is a wide-field multi-wavelength survey of X-ray selected clusters at $z=$0.04-0.07.  The original $34\arcmin \times 34 \arcmin$ WINGS field-of-view has now been extended to cover a 1 deg$^2$ field with both photometry and spectroscopy.
} 
  {In this paper we present the Johnson $B$ and $V$-band OmegaCAM/VST observations of 46 WINGS clusters, together with the data reduction,  data quality and Sextractor photometric catalogs.}
   {The data reduction  was carried out with a modified version of the  ESO-MVM (a.k.a. "ALAMBIC") reduction package, adding a cross-talk correction, the gain harmonisation and a control procedure for problematic CCDs. The stray-light component has been corrected by employing our own observations of populated stellar fields.
}
   {With a median seeing of $1\arcsec$ in both bands, our 25-minutes exposures in each band typically reach the 50\% completeness level  at $V=23.1$ mag.
The quality of the astrometric and photometric accuracy has been verified by comparison with the 2MASS as well as with SDSS astrometry, and SDSS and previous WINGS imaging. Star/galaxy separation and sky-subtraction procedure have been tested comparing with previous WINGS data.
}
   {The Sextractor photometric catalogues are publicly available at the CDS, and will be included in the next release of the WINGS database on the VO together with the OmegaCAM reduced images.
These data form the basis for a large ongoing spectroscopic campaign with AAOmega/AAT and is being employed for a variety of studies.
}

   \maketitle
\section{Introduction}

Galaxy clusters, the most massive collapsed structures in the
Universe, play an important role for both cosmology and galaxy
evolution studies. They are the tail of a continuum distribution of
halo masses, and the most extreme environments where galaxy formation
has proceeded at an accelerated rate compared to the rest of the
Universe. Clusters have been a testbed for studies of galaxy formation
and evolution, uncovering trends that several years later have been
found also in the field \citep{butc+1978, couc+1987, dres+1997}. They are a repository for galaxies that
have been shaped in lower halo-mass environments \citep{wilm+2009}, but also
the sites where essentially all environmental effects are thought to
take place, from strangulation to ram pressure stripping, and even
mergers \citep{delu+2010}.  As peaks in the matter distribution, they
host those galaxies that have formed first and in the most extreme
primordial conditions, and at the same time where the hierarchical
growth is most evident, like brightest cluster galaxies. There is no
better place than rich clusters in the low-z universe to find and
study the descendants of the most massive galaxies observed at high-z \citep{pogg+2013}.

The WIde-field Nearby Galaxy-cluster Survey (WINGS) is a wide-field
and multiwavelength survey of 76 galaxy clusters in the local Universe
\citep{fasa+2006}.  The sample consists of all clusters at
$0.04<z<0.07$ in both hemispheres at Galactic latitude $|b| \geq 20$
selected from the ROSAT X-ray-Brightest Abell-type Cluster Sample, the
Brightest Cluster Sample and its extension \citep{ebel+1996, ebel+1998,
ebel+2000}.

The original WINGS survey is based on $B$ and $V$ imaging for the 76
clusters over a $34 \arcmin \times34'$ field-of-view (FOV) taken with the Wide
Field Cameras on the INT and the 2.2MPG/ESO telescopes \citep{vare+2009}. $J$- and $K$-band Wide Field Camera imaging at UKIRT 
 \citep{vale+2009} and $U$-band with the INT, LBT and
BOK telescopes \citep{omiz+2014} were secured for a subset of
clusters.  Spectroscopy was obtained over the $34\arcmin \times34 \arcmin$ FOV with
2dF/AAT ($\sim4000$ usable spectra) and WYFFOS/WHT ($\sim2500$ spectra) \citep{cava+2009}.
These data allowed us to  derive galaxy morphologies \citep{fasa+2012},
surface photometry and sizes \citep{dono+2014},
stellar masses, star formation histories, and spectral types 
\citep{frit+2011, frit+2014,  vulc+2011}, as well as characterise the
cluster substructure and dynamics \citep[][Cava et al. submitted]{rame+2007}, and conduct a number of studies on galaxy
properties and evolution (a full
publication list can be found at the WINGS website\footnote{\url{http://web.oapd.inaf.it/wings}}).
The WINGS data and data products are publicly available through the
Virtual Observatory (VO), as explained in \cite{more+2014}.

The WINGS optical images, together with the photometry
  and source classification,
were used to calibrate the photometry presented in this paper
and for other purposes, as described in the following sections.

The WINGS dataset is unique, as none of the other low-z surveys
investigate a large sample of clusters and cluster galaxies in such
detail. GAMA offers an exquisite sampling down to low-mass haloes, but
it lacks a large number of massive clusters at redshifts comparable to
WINGS \citep{robo+2011}. The SDSS \citep{york+2000} provides
large cluster catalogues, but has a much lower imaging quality (for
seeing, depth, pixel scale, see below), and is 1.5 magnitudes
shallower than WINGS spectroscopy, yielding a smaller dynamic range of
galaxy magnitudes and masses at the WINGS redshifts.

The main limitation of the original WINGS data is that they cover only
the cluster cores: the maximum clustercentric distance reached in
(almost) all clusters by the INT+2.2m imaging is only 0.6 times the virial radius. 
Crucially, what was missing is the coverage out to at least
the virial radius and into the outer regions. This would be of primary importance, as it would link clusters with
the surrounding populations and the field.

Clusters accrete individual galaxies and larger subclumps from their
outskirts. The outer regions of clusters are the transition regions
between the cores, with their dense and hot intracluster medium,
and the filaments (and/or groups) feeding the cluster, at the point where
galaxies are subject to a dramatic change of environment. Indeed,
observations have proved that the cluster outskirts are essential to
understand galaxy transformations \citep{lewi+2002, pimb+2002,
treu+2003, mora+2007}. Moreover, the projected clustercentric
radius of galaxies statistically retains memory of the epoch when the galaxy
first became part of a massive structure and became a satellite
\citep{smit+2012, delu+2012}. Cosmological hydrodynamical
simulations predict a depletion of both hot and cold gas and a decline
in the star-forming fraction of galaxies as far out as 5 cluster virial radii
\citep{bahe+2013}.  With the exception of a few single clusters and
superclusters \citep[e.g.]{merl+2010, merl+2015, maha+2011, smit+2012,
hain+2011}, this very important transition region between clusters
and the surrounding field remains largely unexplored.

With the aim to cover the virial region and extend out into the infall
region, we have obtained GTO OmegaCAM/VST imaging in the $u$, $B$ and
$V$ bands over $1\times1 deg^2$ for 45 fields covering 46 WINGS clusters. 
A large spectroscopic campaign to follow up
the clusters observed with OmegaCAM is ongoing with AAOmega/AAT
(Moretti et al. in prep.). This imaging+spectroscopic dataset is
from now on named OmegaWINGS.
 
This paper presents the OmegaCAM/VST $B$ and $V$ imaging, the
observations (Sect. \ref{sec:obs}), data reduction (Sect. \ref{sec:datared}), the release of
photometric catalogs (Sect. \ref{sec:cata}) and data quality (Sect. \ref{sec:qc}).  The $u$-band
campaign is still ongoing and will be presented in a subsequent paper.

In the following, we use $H_0=70 \, \rm km \, s^{-1} \, Mpc^{-1}$,
$\Omega_m=0.3$, $\Omega_{lambda}=0.7$.

\section{Observations}\label{sec:obs}

The VLT Survey Telescope \citep[VST,][]{capa+2011}
is a 2.6-m wide field optical telescope placed at Cerro Paranal, Chile.
The telescope is equipped with OmegaCAM \citep{kuij+2011}, 
a camera that samples the 
1 deg$^2$ VST unvignetted field of view  with a mosaic of
32 4k$\times$ 2k CCDs at $0\farcs21$/pix.
The layout of the OmegaCAM mosaic is shown in Fig. \ref{f:flat}; the ESO identification name is superimposed to
each CCD.
\myfigure{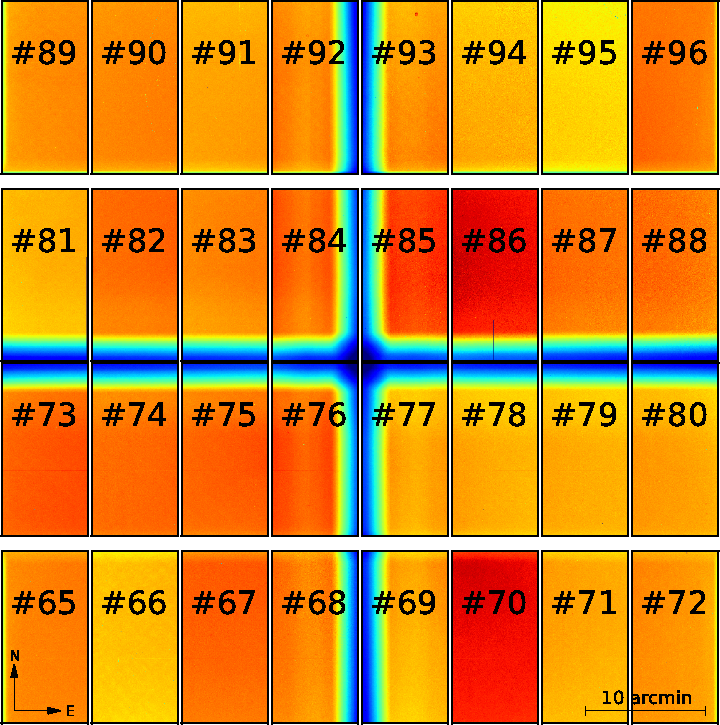}{
Layout of the OmegaCAM CCD mosaic.
The labels indicate the ESO ID of each chip.
The image is a $V$-band raw flat field image. 
}{f:flat}

\myfigure{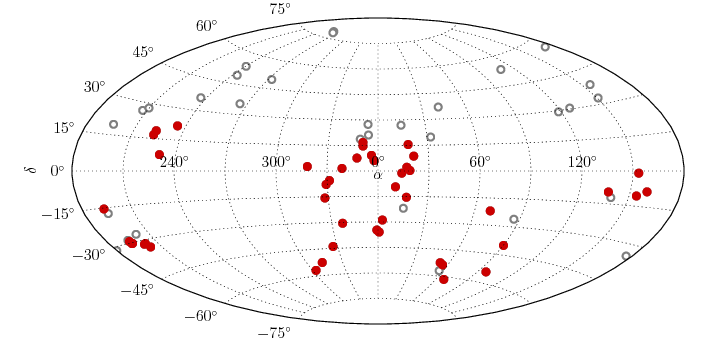}{OmegaWINGS target clusters are shown as 
filled circles, all WINGS clusters are shown as open circles.}{f:targets}

OmegaWINGS target clusters 
were randomly selected from the
57 WINGS  clusters that can be observed from VST ($\delta<20^\circ$).
We obtained service-mode $B$ and $V$-band imaging for 46 of them with 45 OmegaCAM
pointings.
Two WINGS clusters --A3528a and A3528b-- were observed with a single VST pointing;
hereafter this will be referred to as A3528.
The position of the target clusters observed by the OmegaWINGS survey are shown  in
Fig. \ref{f:targets}. 
The choice of $B$ $V$ filters was taken for consistency with the original WINGS survey, despite the problems related with the segmentation of OmegaCAM Johnson's filters
that are discussed at the end of this section.

Observations started in October 2011 and were concluded in September 2013.
The first observations were carried out during ESO period P88 with the OmegaCAM {\it STARE-mode},
splitting the total exposure time in $3\times 480s$ observations with no offsets.
We adopted this observing mode to obtain a constant signal to noise ratio across the field of view,
as in the original WINGS survey.  
Starting from period P89 we optimised our observing strategy, taking  $5\times 300s$ exposures in {\it DITHER-mode},
with $25\arcsec$ and $85\arcsec$ offsets in horizontal and vertical direction, respectively. This observing strategy offers two major advantages: 
it allows to dither out the gaps between the CCDs and to estimate the contribution of the background light (see next section for details).
The log of our observations is summarised in Table \ref{tab:log}.

On average, observing conditions were better in $V$-band than in $B$-band.
We measured the seeing on each OmegaWINGS image  as the mean value of
the FWHM of stars profiles. The values are listed in Table \ref{tab:log} and shown in
Fig. \ref{f:seeing}. 
The seeing is lower than $1\farcs3$ in 80\% of $B$-band images,
and lower than $1\farcs2$
in 80\% of $V$-band ones.
The median values of seeing are $1\farcs0$  in both $B$- and
$V$-band.

Since OmegaCAM $B$ and $V$ filters
are segmented and composed by 4 quadrants, the interface of the
quadrants casts a slight shadow in the form of a cross onto the image
plane. This central vignetting cross is $\sim310\arcsec$ wide in both directions.  
Figure \ref{f:flat} shows a raw flat-field image in the $V$-band, where the vignetting is clearly visible.
To
remove it from the final stacked images, the OmegaCAM User
Manual\footnote{
  \url{http://www.eso.org/sci/facilities/paranal/instruments/omegacam/doc.html}}
suggests a dithering pattern with steps that should be $310\arcsec$ in
both $X$ and $Y$ direction. Such wide steps would reduce the 
field of view covered by 5 exposures. Considering that the central region
of each target cluster is covered by WINGS data, we 
decided to use smaller dithering steps.
However, in this
way the vignetting cross can not be entirely removed, and our final
images have a vertical stripe $\sim 3 \arcmin$ wide that is strongly
affected by vignetting and that was therefore masked out.  The
horizontal component of the vignetting cross was instead perfectly removed.

\myfigure{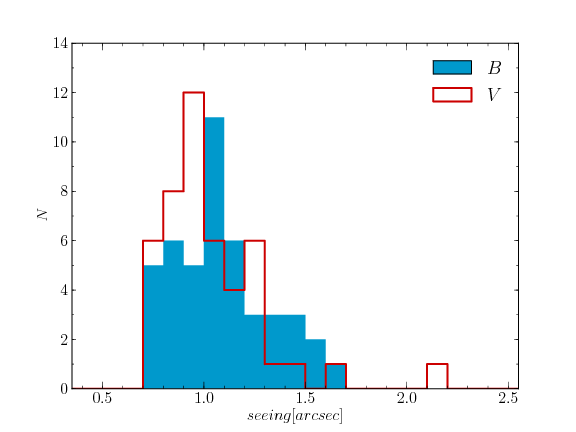}{Mean seeing measured on stacked OmegaWINGS $B$- and $V$-band images
}{f:seeing}

\mytable{l ll ll l}{
  cluster & 
 DATEOBS$_B$ &$\sigma_B$&
 DATEOBS$_V$ &$\sigma_V$&
src}{
A85           &2013-09-03 &  $0\farcs97$ &2013-08-03 &  $1\farcs00$&  SDSS \\ 
A119          &2011-12-17 &  $1\farcs03$ &2011-10-23 &  $0\farcs74$&  SDSS \\ 
A147          &2013-07-15 &  $0\farcs78$ &2013-08-05 &  $0\farcs83$&  SDSS \\ 
A151          &2012-11-17 &  $0\farcs85$ &2012-11-04 &  $0\farcs75$& 2MASS \\ 
A160          &2011-10-21 &  $0\farcs79$ &2011-10-21 &  $0\farcs99$&  SDSS \\ 
A168          &2013-07-18 &  $1\farcs17$ &2013-08-03 &  $1\farcs23$&  SDSS \\ 
A193          &2011-10-21 &  $0\farcs78$ &2011-10-21 &  $1\farcs01$&  SDSS \\ 
A500          &2011-11-28 &  $1\farcs26$ &2011-12-02 &  $1\farcs28$& 2MASS \\ 
A754          &2011-11-30 &  $0\farcs76$ &2011-11-22 &  $0\farcs95$& 2MASS \\ 
A957x         &2012-03-26 &  $1\farcs05$ &2011-11-23 &  $1\farcs02$&  SDSS \\ 
A970          &2011-12-23 &  $1\farcs64$ &2011-11-24 &  $1\farcs25$& 2MASS \\ 
A1069         &2013-04-13 &  $1\farcs31$ &2013-05-07 &  $0\farcs87$& 2MASS \\ 
A1631a        &2013-03-22 &  $1\farcs16$ &2013-02-10 &  $0\farcs98$& 2MASS \\ 
A1983         &2012-05-18 &  $1\farcs05$ &2012-03-31 &  $1\farcs23$&  SDSS \\ 
A1991         &2013-04-15 &  $0\farcs86$ &2013-04-14 &  $0\farcs84$&  SDSS \\ 
A2107         &2013-04-08 &  $1\farcs03$ &2013-04-10 &  $1\farcs01$&  SDSS \\ 
A2382         &2012-07-20 &  $1\farcs03$ &2012-06-26 &  $2\farcs12$& 2MASS \\ 
A2399         &2012-06-16 &  $0\farcs84$ &2012-05-29 &  $1\farcs24$&  SDSS \\ 
A2415         &2012-07-26 &  $1\farcs49$ &2012-07-22 &  $0\farcs82$&  SDSS \\ 
A2457         &2012-06-16 &  $1\farcs08$ &2012-07-15 &  $1\farcs13$&  SDSS \\ 
A2589         &2013-07-16 &  $1\farcs22$ &2013-07-13 &  $0\farcs96$&  SDSS \\ 
A2593         &2012-10-08 &  $1\farcs41$ &2012-10-08 &  $1\farcs01$&  SDSS \\ 
A2657         &2013-07-17 &  $0\farcs78$ &2013-07-11 &  $0\farcs77$&  SDSS \\ 
A2665         &2013-07-12 &  $0\farcs96$ &2013-07-12 &  $0\farcs96$&  SDSS \\ 
A2717         &2013-08-01 &  $1\farcs57$ &2013-06-11 &  $1\farcs22$& 2MASS \\ 
A2734         &2013-06-20 &  $1\farcs13$ &2013-07-07 &  $1\farcs06$& 2MASS \\ 
A3128         &2011-12-20 &  $1\farcs03$ &2011-12-18 &  $0\farcs77$& 2MASS \\ 
A3158         &2011-12-18 &  $0\farcs95$ &2011-12-20 &  $0\farcs93$& 2MASS \\ 
A3266         &2012-10-15 &  $1\farcs53$ &2012-10-15 &  $1\farcs10$& 2MASS \\ 
A3376         &2013-01-04 &  $1\farcs01$ &2012-11-17 &  $1\farcs32$& 2MASS \\ 
A3395         &2013-03-05 &  $0\farcs89$ &2013-03-02 &  $1\farcs11$& 2MASS \\ 
A3528         &2013-06-02 &  $1\farcs43$ &2013-06-05 &  $1\farcs11$& 2MASS \\ 
A3530         &2013-06-03 &  $0\farcs95$ &2013-06-06 &  $0\farcs86$& 2MASS \\ 
A3532         &2013-06-03 &  $0\farcs91$ &2013-06-07 &  $0\farcs77$& 2MASS \\ 
A3556         &2012-06-17 &  $1\farcs21$ &2012-05-24 &  $1\farcs44$& 2MASS \\ 
A3558         &2013-06-11 &  $0\farcs85$ &2013-06-28 &  $0\farcs76$& 2MASS \\ 
A3560         &2012-06-18 &  $0\farcs89$ &2012-05-24 &  $1\farcs68$& 2MASS \\ 
A3667         &2013-04-13 &  $1\farcs38$ &2013-05-14 &  $0\farcs95$& 2MASS \\ 
A3716         &2013-05-20 &  $1\farcs13$ &2013-05-20 &  $0\farcs93$& 2MASS \\ 
A3809         &2012-07-22 &  $1\farcs12$ &2012-04-18 &  $0\farcs99$& 2MASS \\ 
A3880         &2013-06-11 &  $1\farcs31$ &2013-06-20 &  $0\farcs92$& 2MASS \\ 
A4059         &2013-08-04 &  $1\farcs05$ &2013-07-03 &  $0\farcs91$& 2MASS \\ 
IIZW108       &2013-06-06 &  $1\farcs04$ &2013-06-06 &  $0\farcs86$&  SDSS \\ 
MKW3s         &2012-04-20 &  $1\farcs14$ &2012-04-19 &  $0\farcs83$&  SDSS \\ 
Z8852         &2012-11-10 &  $1\farcs02$ &2012-10-12 &  $0\farcs83$&  SDSS \\ 

}{Observation log. Columns 3 and 5 are seeing in $B$ and $V$ measured as the average FWHM of stars
in $B$- and $V$-band final stacked images. The last column lists the reference astrometric catalogue.}{tab:log}

\section{Data reduction}\label{sec:datared}

\begin{figure}
  \resizebox{\hsize}{!}{\includegraphics{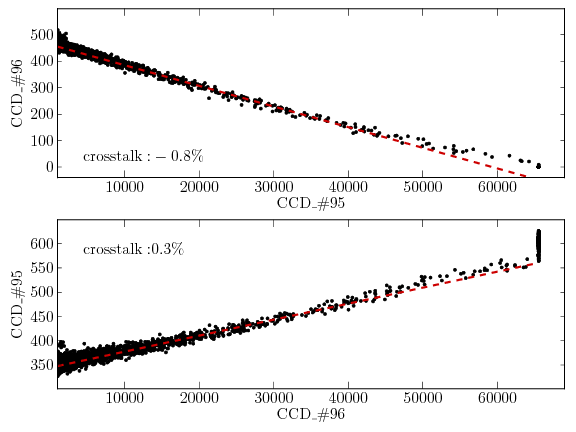}}
  \caption{%
  Cross-talk effect between CCDs \#95 and \#96.
  Each point is the
  count in one pixel in the raw frame of the {\it receiver} CDD as a
  function of the count in the same pixel in the {\it emitter}.  
  There  is a positive cross-talk from \#96 to \#95 ($\sim 0.3\%$) and a 
  negative one from \#95 to \#96 ($\sim -0.8\%$). Only data-points corresponding
  to  more than $2500$ ADUs in the {\it emitter} has been plotted.
}
\label{f:xtalk}
\end{figure}

\begin{figure}
  \resizebox{\hsize}{!}{\includegraphics{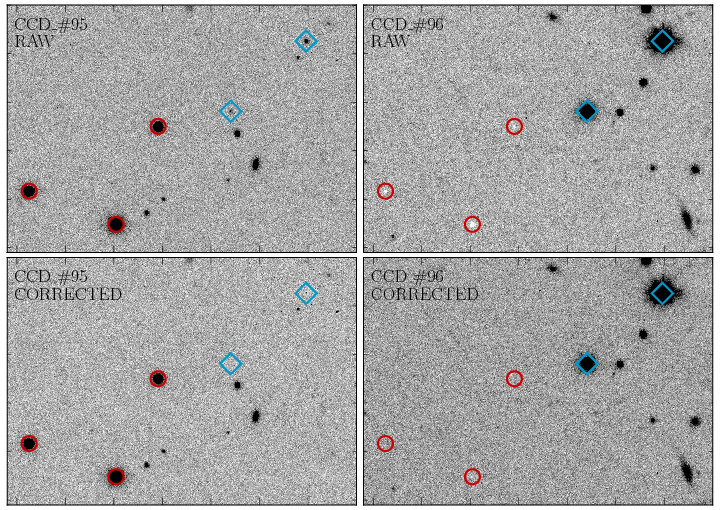}}
  \caption{%
  Cross-talk effect between CCDs \#95 and \#96.
  The top panels show how bright sources in CCD \#96 produce 
  positive signal on CCD \#95, while bright sources in CCD \#95 generate 
  ``holes'' on CCD \#96.
  The corrected images are shown in the {\it
  bottom panels}.
  Circles and diamonds mark the position of bright sources on  CCDs \#95 and \#96, respectively.
  Fake sources are removed from the image in the lower-left panel, as well as
  holes from the lower-right one.
}
\label{f:xtalkima}
\end{figure}

Image reduction and calibration are mainly based on ESO-MVM -also known
as {\sc alambic}- reduction package.  This is a multi-instrument
reduction tool originally developed for the ESO/EIS survey
\citep{mign+2007}.  It has been extensively used also in the
production of ESO Advanced Data Products, see for instance the 30
Doradus/WFI Data
Release\footnote{\url{http://archive.eso.org/archive/adp/ADP/30_Doradus/}},
or the GOODS/ISAAC Final Data Release \citep{retz+2010}.
The detailed description of the package and the documentation of the
algorithm structure implemented in ESO/MVM are given in
\cite{vand2004}. This section presents a summary of the main reduction steps
and our add-ons to the original pipeline. 
The latest version of the {\sc alambic} code and User
Manual can be downloaded at the ESO webpage\footnote{
  \url{http://archive.eso.org/cms/eso-data/data-packages/eso-mvm-software-package.html}}.
Here we used a modified version of the code \citep[kindly provided by H. Bouy and B. Vandame, see][]{bouy+2015}
that has been partially rewritten to take advantages of  the most recent hardware
and recent Linux libraries.
There are configuration files for many optical and near-infrared ESO
instruments, but OmegaCAM is not officially supported so far. We
therefore created a new configuration file for OmegaCAM, using the
instrument description given in the VST user manual.

The following subsections will describe the main reduction
steps. The only steps for which we had to develop integrations to 
{\sc alambic} are:
the cross-talk correction, the gain-harmonisation and the control
procedure to check the quality of CCD \#82.

We will call Data Block (DB) the complete set of data
taken in each photometric band for a single pointing. A DB consists of
5 science, 5 twilight flat-field and 10 bias frames.
Each DB has been reduced independently. This may slightly increase
the computational time, because some targets have
been observed during the same night and therefore it would have been possible
to compute the master bias and flat-field frames only once. However,
we prefer to reduce each DB 
independently
because in this way the
implementation of the reduction pipeline is much easier and linear.
The calibration stacking process is not very time
consuming, therefore our choice has a negligible influence
on the overall computational efficiency of the reduction process.
A typical reduction run for a DB
takes about 40 minutes on a Intel i7 3.4GHz computer with
16 Gb of RAM. 

\subsection{Data organisation}
First of all the multi-extension raw image files are splitted,
resulting in 32 single-extension files, corresponding to the 32
OmegaCAM detectors. 
Images are then classified and grouped together using 
the information stored in the file headers.
{\sc alambic} creates 
lists of images corresponding
to consecutive observations of the same field taken with the same
filter, called {\it observation blocks}.
These are used to produce the
{\it calibration blocks}, i.e. lists used to create the calibration images
--bias, flats, illumination maps--
and the
{\it reduction blocks}, i.e. lists of science observations
of each scientific target with the same filter.

\subsection{Cross-talk}

According to the OmegaCAM User Manual, four detectors (CCDs \#93-96, see Fig. \ref{f:flat}) suffer
electronic cross-talk. The strongest effect, of the order of a few percents, is between CCDs \#95 and
\#96, while it is much lower for all the other CCDs.  
After a few tests we confirmed the cross-talk for 
CCDs \#95 and \#96 and we found that it is negligible in all other cases.
The cross talk has
been estimated by cross-correlating the signal registered on the same
pixel in each pair of CCDs. As an example, Fig. \ref{f:xtalk}  shows the
effect in
a raw image of one of our science frames. 
The mean background level for
this image is $\sim 480$ ADUs in CCD \#96 and $\sim 350$ ADUs in CCD
\#95. When a bright source increases the signal in CCD \#96, a signal
above the background is registered in CCD \#95 (see lowest panel in
Fig. \ref{f:xtalk}); the difference between the registered signal
and the average background in one detector is proportional to the
signal in the other one.  Figure \ref{f:xtalk} shows deviations from the
linear relation when the signal is above $\sim 50000$ ADUs. This is
mainly due to non-linearity of the detectors. This non-linearity can be safely
ignored, since it affects only a few pixels in each image. We also
note that the cross-talk effect of saturated stars in CCD \#95
inversely saturates  CCD \#96 at 0 ADU. To avoid this problem, on
September $12^{th}$ 2012 the bias level of CCD \#96 was increased to
650 ADU.

Since {\sc alambic} does not include any cross-talk analysis, we developed a fast and easy procedure to calculate and correct for cross-talk.
We assumed that the {\it observed} image
is equal to
\begin{equation}\label{eq:xtalk}
S_{r}=I_{r}+\alpha^{e}_{r} \cdot S^{e}
\end{equation}
where $S_r$ and $I_r$ are the observed and the {\it real} --i.e. if no
cross-talk were present-- signal in the receiver, and $S^e$ is the
observed image in the emitter detector. $\alpha^{e}_{r}$ is the
cross-talk coefficient between the receiver and the emitter CCDs.  The
coefficient has been obtained by fitting a linear relations to the
data-points in Fig. \ref{f:xtalk}. In all our images the
coefficients for CCDs \#95 and \#96 are very similar,
$\alpha^{95}_{96}\simeq -0.8\%$ and $\alpha^{96}_{95}\simeq 0.3\%$.
These values are then used to correct the images, inverting
eq. \ref{eq:xtalk}.  
Figure \ref{f:xtalkima}  shows the very good results obtained with our procedure.

\subsection{Stacking of calibration frames}
During this phase of the reduction process, all calibration images of a given
science reduction block 
are 
stacked together and the corresponding
master bias and twilight flat images are produced. During this step,
 a bad-pixel detection procedure is applied
and weight maps for flat field images are computed. Details about the 
algorithm  are given in the {\sc alambic} User Manual.

\subsection{Gain harmonisation}\label{s:gain}

The electronic converters of each detector are different, and each CCD
may have a different efficiency. Therefore, each detector has its own
effective gain and, as a consequence, its own photometric zero-point.
The chip-to-chip gain variation quoted in the OmegaCAM User Manual is
of the order of $10\%$, resulting in a chip-to-chip zero-point scatter of
 $\sim 0.1$ mag.
The procedure adopted by {\sc alambic} to correct for this is to
apply a multiplicative calibration constant to the master flats.
The calibration is based on the analysis of a scientific image.  The
chip background is computed for each of the four borders of each chip
in a narrow stripe. Then the chip-to-chip gain variations are
calculated by comparing the values of each pair of adjacent stripes on
different chips.
As an example, in the case of a camera with $4\times2$
CCDs --as the WFI@ESO2.2m instrument--, the total number of equations,
one for each pair of stripes, is 10, and the unknown parameters are
the seven unknown flux-scales (this is a relative calibration).
In this way it is possible to obtain a robust calibration even if
the sky-background is not constant and presents, for example, some gradients.

The central vignetting cross does not allow to use the standard {\sc alambic} procedure for gain harmonisation.
Given the high flux loss in a wide cross-shaped
region at the centre of the field of view of the camera  it is not
possible to easily connect the background values of adjacent regions
of different CCDs affected by the central vignetting.  We
therefore developed a variation of the original {\sc alambic}
procedure optimised for our specific OmegaCAM observations.
First of all we
note that the background level in our raw images
does not show any significant gradient within each single
detector. We can therefore assume a constant sky background
across the whole FoV.  We took one of the detectors as a reference and
scaled the master flat field image of the other 31 according to the ratio of
the mean background values for each CCD on a science image.  
We recall that the reduction process is done independently
for each DB, and hence also this process is repeated for
each OmegaWINGS field in each filter. This procedure can be used since all our
science images are not extremely crowded. The background estimation,
and consequently the gain-harmonisation, would not have been possible
otherwise, as in the case of observations of e.g. giant nearby galaxies
with sizes of the order of one OmegaCAM CCD, or observations
of the central regions of galactic globular clusters, nearby resolved galaxies
 or other crowded stellar fields. In such cases it would be difficult --if not impossible--
to estimate the sky background in each detector.

When the re-scaled master flat-field images are used to calibrate scientific
images, the chip-to-chip gain variation is corrected and the resulting calibrated
images  therefore have a uniform background value. The quality of
this procedure will be discussed in the following sections, in the
context of the discussion of the overall photometric performances of our
reduction pipeline.

To conclude this section, we add an important note about CCD \#82 (its location on OmegaCAM mosaic is shown in  Fig. \ref{f:flat}). The
OmegaCAM User Manual report day-to-day gain variations of a few
percent, since the start of OmegaCAM operations. We note that for many
observations there are serious problems on CCD \#82. The background
value is not constant, showing strong discontinuities in the form of
horizontal stripes with different background values. The position and
the extension of these stripes are in general different from one image
to the other, even for consecutive observations within the same DB.
For this reason, we decided to discard all CCD \#82 data when one scientific image in the DB is
affected by this problem.
  We implemented in the pipeline a simple script  to test whether
CCD \#82 background is stable. If this is not the case, all CCD's
pixels are assigned a null weight.
On June $2^{nd}$ 2012 ESO changed the video board connected to CCD \#82 to
fix this issue.

\subsection{Illumination correction}\label{sec:illcorr}

A well known problem affecting in particular wide field cameras is the
sky-concentration, i.e. stray-light component centrally
concentrated. This mostly affects observations with extremely bright
background. In particular, flat-field frames are the most exposed to
this effect. The net result, when flat-field exposures are applied
without any correction, is an erroneous apparent trend of photometric
zero point with distance from the centre of the camera field of view. 

A common technique
used to correct for this effect is to compute an illumination map to be
applied to the flat-field frames, in order to obtain {\it
  photometrically flat} reduced science frames. {\sc alambic} implements
an algorithm to compute and apply the illumination map, which is based
on a sequence of dithered observations of the same stellar field.
Basically, these images are reduced using the {\it normal} flat field
and photometric catalogs are extracted for each image.  Since each
star will be placed at a different position in each image, it is
possible to map the zero-point variations as a function of the
position on the focal plane. This is done using a least square
estimator with rejection approximating the illumination map with a 2D
polynomial function. This map is finally multiplied by the flat-field
frame.  A detailed description of the algorithms is given in
\cite{vand2002}.

\myfigure{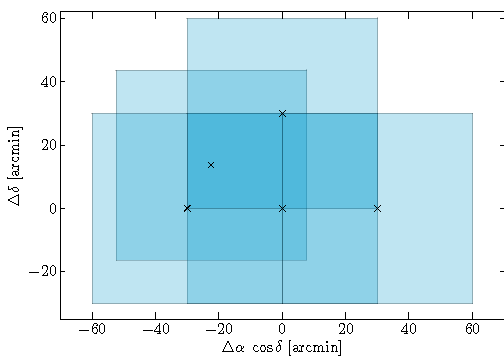}{Dithering pattern used for the
  observation of the SA107, which has been used to compute the
  illumination map.  {\it Crosses} mark the five centres  of the
  pointings, while the {\it rectangles} represent the OmegaCAM FoV.
}{f:illmapobs}

To compute the illumination map we observed a well populated stellar
field, namely the Landolt SA107. We obtained 5 images in both the $B-$
and $V$-bands, using a dithering path wide enough to obtain observations of
the same stars in different positions in the OmegaCAM focal plane.
An outline of the dithering pattern is shown in 
Fig. \ref{f:illmapobs}.
These observations were used to compute the illumination map, using
a 4th order polynomial function. This illumination-correction map has then been applied to
the flat-field frames used to calibrate all science frames.

In wide-field instruments, the illumination variation pattern across
the large detector block can vary with time, telescope position,
etc. 
The 
OmegaCAM consortium
reported a dependence of the OmegaCAM illumination map on the telescope
rotator angle\footnote{\url{http://www.eso.org/sci/facilities/paranal/instruments/omegacam/doc/OCAM_illum.pdf}}.
 It has been however pointed out that to achieve a photometric
accuracy at the 1\% level, the illumination map can be considered
{``\it stable on a timescale of at least 7 months"}.  
Our illumination map was computed from observations taken on July 2012, 
1 year after the first OmegaWING observation and  1 year before the last one.
To check the stability of our illumination correction, we compared OmegaWINGS
photometry with WINGS and SDSS one and found no relevant variation of the photometric
zero point across our images (see Sect. \ref{sec:qc}).

\subsection{Stacking of science frames}\label{sect:scistack}
\begin{figure*}
  \resizebox{\hsize}{!}{\includegraphics{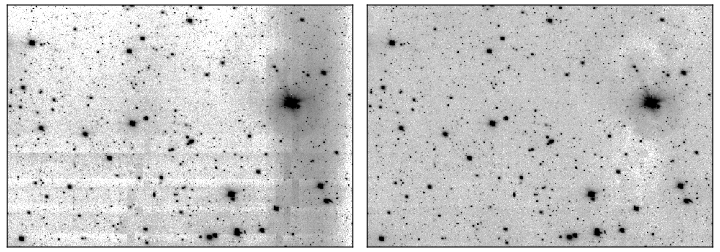}}
  \caption{
Zoom of a $5000\times3500$ pixel region of two $V$-band stacked images of A2415 obtained using different procedures.
The sky background has not been subtracted on the image in the 
{\it left panel}.  The final product of our pipeline is shown 
in the {\it right panel}.  
}
  \label{f:stack}
\end{figure*}

During this stage the pipeline finally operates on the science frames,
using the calibration frames obtained from the previous steps.  As 
part of this reduction stage, our pipeline computes and subtracts from
the images the additive sky-background contribution.  This must be
done since there are stray-light components mainly due to reflections
due to the segmented filters.  This effect can be seen
in the left panel of Fig. \ref{f:stack}, where a $V$-band image obtained stacking all observations of
A2415 is shown.
Only bias-subtraction and
flat-field correction has been applied to these images, which are then
stacked together without any further processing. For the sake of
clarity, only a $5000\times3500$ pixels region is displayed. It nearly
corresponds to the upper-left quadrant of the  mosaic (CCDs \#82,\#83,\#84,\#90,\#91,\#92, see Fig. \ref{f:flat})
An excess of light due to
light scattered by the filter support
is clearly visible on the right side of the image on the left panel.
In addition, in the lower half of the images there are
small discontinuities. These are the footprints of the borders of
individual detectors on the five stacked images.  These
discontinuities are likely due to a non perfect flat-field correction.
It is worth noticing, however, that the image in Fig. \ref{f:stack} is displayed
using a power-law gamma correction that strongly
enhances the low-brightness details.  These discontinuities are of the
order of the standard deviation of the background signal. 
If these discontinuities were due to small uncertainties in the flat
fielding, they should have been corrected as a multiplicative component.
This is not the case; in fact our sky-subtraction procedure eliminates them from the final
stacks (as can be seen in the right panel of Fig. \ref{f:stack}).
This means that they are considered as additive contribution.
This possible mis-interpretation would introduce a minor
bias in the photometric zero-point in the regions of the mosaic
corresponding to the CCDs borders.
Considering
their limited extension, and the fact that they are present only in a
few images --the reason for this is not completely clear--, they can
be ignored, as they will not affect the quality of the photometric
calibration at levels higher that a few percent, that is the
requirement for our scientific programme.

The presence of the additive stray-light component would not be an
issue for stellar photometry, but we must correct for it since we are
interested in surface photometry of extended sources.
First of all, a standard calibration of the science frames is
performed. Over- and pre-scan regions are trimmed from raw images,
these are then bias-subtracted and flat-field corrected.

Then, the sky-background is computed assuming that it is constant for the
five consecutive images belonging to the same DB. Under this
assumption it can be computed with an algorithm similar to the one
commonly used to remove the background from infrared observations or to
correct for fringing patterns; indeed we used the {\sc alambic} fringing map
estimator,
fully described in the {\sc alambic} User Manual. Further
details of the algorithms are also given by \cite{vand2004}.

The astrometric calibration is performed for all frames using as 
reference the 2MASS catalog \citep{skru+2006} or the SDSS DR8
\citep{sdss8}, when available.  Astrometric distortions are mapped
using a polynomial function of order four.  The absolute accuracy
--measured on the final stacked mosaic-- is of the order of $0\farcs2$
and $0\farcs07$ when the calibration is based on 2MASS and SDSS,
respectively.

Satellite tracks are detected using a Hough-transform algorithm to
search for straight lines in raw images. These are masked and flagged
as bad pixels.

All images are then warped using the astrometric solution and
projected in an user-defined common grid. We defined a distortion-free
grid with a constant pixel scale equal to the average OmegaCAM pixels
scale, i.e $0\farcs213$. The grid is centred at the target cluster
centre. All warped images are finally stacked together, using the
single weight maps. The output is the final stacked mosaic and the
corresponding weight map. We note that since we used the same
reference grid for $B$- and $V$-band images, the two output mosaics
for each cluster are already aligned.

\section{Photometric catalogs}\label{sec:cata}

\subsection{Source extraction}
The source extraction and measure of photometric and structural
parameters has been performed using {\sc sextractor} \citep{bert+1996}.
$B$-band photometry has been carried out with Sextractor {\it dual-mode}, using $V$-band image as reference.
The catalogues extracted from $B$- and $V$-band images were then matched together using a searching radius of $2\arcsec$.
In the following we list the parameters we measured. For a detailed 
description of the algorithm used to derive them we refer to Sextractor User Manual\footnote{\url{http://www.astromatic.net/software/sextractor}}.
\begin{description}
\item[\texttt{RA}, \texttt{DEC}:] equatorial coordinates of the barycentre of the source emission profile;
\item[\texttt{RA\_PEAK}, \texttt{DEC\_PEAK}:] equatorial coordinates of the source emission peak;
\item[\texttt{X}, \texttt{Y},\texttt{X\_PEAK}, \texttt{Y\_PEAK}:] coordinates on the image, in pixels, of source barycentre and emission peak;
\item[\texttt{ISOAREA\_IMAGE}:] isophotal area;
\item[\texttt{KRON\_RADIUS}:] Kron radius;
\item[\texttt{FWHM\_IMAGE}:] full width at half maximum;
\item[\texttt{A\_IMAGE}, \texttt{B\_IMAGE}:] semi-major and semi-minor axes. This has been used to compute the axial ratio $b/a$;
\item[\texttt{THETA\_IMAGE}:] position angle with respect to the North and measured counter-clockwise;
\item[\texttt{CLASS\_STAR}:] stellarity index;
\item[\texttt{MU\_MAX}:] surface brightness of the brightest pixel;
\item[\texttt{MAG\_ISO}:] isophotal magnitude, defined using the detection threshold as the lowest isophote;
\item[\texttt{MAG\_ISOCOR}:] Isophotal magnitude corrected to  retrieve the fraction of flux lost by isophotal magnitudes by assuming Gaussian intensity profiles.
As reported in the Sextractor user manual, {\it "this correction works best with stars; and although it is shown to give tolerably accurate results with most disk galaxies, it fails with ellipticals because of the broader wings of their profiles"};
\item[\texttt{MAG\_AUTO}:] Kron-like aperture magnitude. This is the most precise estimate of total magnitudes for galaxies;
\item[\texttt{MAG\_APER}:] Aperture magnitude. We used apertures with diameter  5, 10, 15, and 20 pixels; $1\farcs60$, $2\farcs00$, and $2\farcs16$;
4, 10 and 20 kpc. To calculate the last three apertures we used the cluster distance
listed in the WINGS database \citep{more+2014}.
\end{description}

\subsection{Photometric calibration}\label{sec:calib}

Photometric calibration has been done using WINGS
stars as local standards. We fitted the equations

\begin{eqnarray} 
  B_{STD} - b &=&a_B \ (B-V)_{STD}+b_B\\
  V_{STD} - v &=&a_V \ (B-V)_{STD}+b_V
\end{eqnarray}

The data have been fitted imposing the condition that the colour-term ($a_B$ and $a_V$)
is constant within each ESO observing semester. As an example, Fig. \ref{f:calib} shows the calibration
relation fitted to MKW3s data.
The results for all OmegaWINGS clusters are listed in Table \ref{tab:calib} in the Appendix.
Colour terms variations are within 0.015 mags.

\myfigure{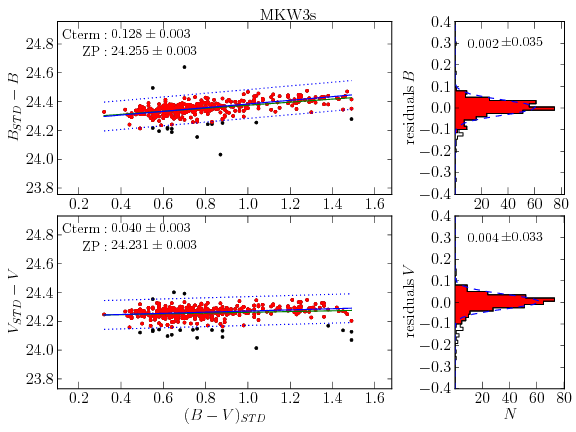}{An example of the photometric calibration fit
  obtained for all stars in common with previous WINGS photometry. In
  this example we used data from MKW3s observations.  {\it Left
    panels} show the difference between calibrated WINGS INT photometry
   and instrumental OmegaCAM magnitudes as a
  function of the $B-V$ colour. {\it Upper and lower panels} show the
  results for $B$- and $V$-band photometry, respectively. The linear fit shown as a {\it blue line}
  was obtained clipping out outliers {\it (black dots)}.  The
  distribution of the residuals is well described by a gaussian
  function, shown in the {\it right panels}.  }{f:calib}

\subsection{Star-galaxy classification}

\begin{figure*}
  \centering
  \resizebox{.8\hsize}{!}{\includegraphics{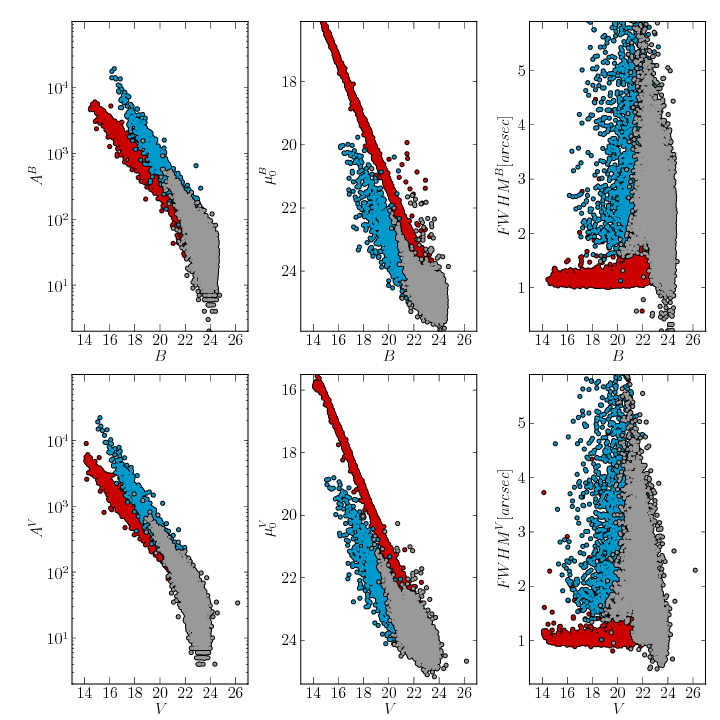}}
  \caption{
Some of the plots used to classify sources in A3809, see text for details.
In each panel stars are shown in red, galaxies in blue and unclassified sources in grey.}
  \label{f:class}
\end{figure*}

The classification of OmegaWINGS sources was done following the method and
criteria used for the original WINGS survey, as described in \cite{vare+2009}.
As a starting point, we classified objects on the basis of 
  the Sextractor \texttt{CLASS\_STAR} parameter:
\begin{description}
\item[stars]: \texttt{CLASS\_STAR}$\geq0.8$
\item[galaxies]: \texttt{CLASS\_STAR}$\leq0.2$
\item[unknown]: $0.8<$\texttt{CLASS\_STAR}$<0.2$
\end{description}
We then used a set of diagnostic plots, using different
combinations of Sextractor parameters to check the result and eventually correct any misclassification.
As an example, 
in Fig. \ref {f:class}  the isophotal area ($A^B, A^V$), the
central surface brightness  $(\mu_0^B, \mu_0^V)$ and the FWHM of sources in 
the A3809 field are plotted as a function of the total magnitude in
both the $B$- and $V$-bands.
Other parameters used for the diagnostic plots include the ellipticity and the difference between
aperture photometry at 5 and 15 pixels.
We visually  checked all clusters looking for outliers in the 
diagnostic plots, i.e. sources mis-classified on the basis of
the automatic classification based on \texttt{CLASS\_STAR}. For some
of them we could safely redefine the star/galaxy classification, after a careful visual inspection
of their $B$- and $V$-band images.
In some cases --faint and/or compact sources-- $B$- and $V$-band photometry provided a different classification; in these cases 
we based our classification on the results provided in the band observed under the best seeing conditions.
For some of the faintest objects, 
with properties between those of stars and galaxies,
the classification remains {\it unknown}.
The reliability  of our classification will be analysed in 
  Sect. \ref{sec:qc}.

\subsection{Data retrieval}
All  {\sc sextractor} measurements for all galaxies are publicly available at CDS
 as a single table; a unique ID is assigned to each
galaxy. To this end we cross-matched the
OmegaWINGS catalogue with the WINGS database \citep{more+2014}.
For galaxies already in the WINGS database we took the WINGS-ID,
while we defined a new ID for all other galaxies.
A list  of the columns of the catalogue is given in Table \ref{tab:tabdesc}. 
The full OmegaWINGS catalogue will be included in the next release of
the WINGS database. This is planned at the end of our AAOmega spectroscopic survey.
We note however that the OmegaWINGS catalogue at CDS, as any other CDS table,
is already part of the VO and can therefore be easily 
cross-matched with the WINGS database using any VO tool, e.g. STILTS.

\section{Data reduction quality checks}\label{sec:qc}
\subsection{Astrometry}

\myfigure{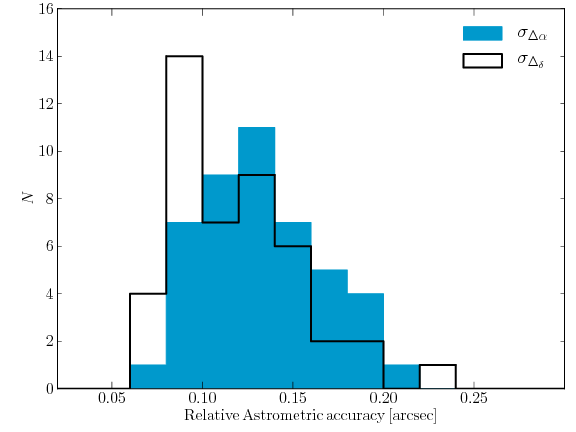}{Dispersion of the distributions of sky-coordinates differences between OmegaWINGS and WINGS
positions of all stars in each 46 OmegaWINGS fields. Only stars brighter than $V=20$ were used. 
  }{f:astrometry}
  
The astrometric accuracy of our catalogs has been
tested against the 2MASS and SDSS DR8 -when available- stellar
catalogs.  By comparing the difference in the source positions, we
verified that no residual distortions are present in the final
mosaics.  The absolute astrometric accuracy is well within the
precision required for the purposes of our scientific project.
For each cluster in our sample, we compared  OmegaWINGS sky coordinates of all stars in the field of view with
2MASS or SDSS (depending on the  catalogue used as the astrometric reference) ones. The distributions of the differences in $\alpha$ and $\delta$ coordinates
 have always negligible mean values, and typical dispersions of
$0\farcs2$ (2MASS) and $0\farcs07$ (SDSS).
As a further tests of the astrometric calibration accuracy, we compared the sky-coordinates of all stars in OmegaWINGS and WINGS catalogs.
The dispersion of the $\Delta\alpha$ and $\Delta\delta$ distributions is a robust indicator of the accuracy of our astrometric calibration, as WINGS was  calibrated
independently. Results are shown in Fig. \ref{f:astrometry}  and confirm that the  astrometry internal calibration is accurate at a level always better than
$0\farcs2$\footnote{The internal accuracy of WINGS astrometry is $\sim0\farcs2$ \citep{fasa+2006}; values in Fig. \ref{f:astrometry} are therefore
upper limits of the  OmegaWINGS astrometric calibration uncertainties. }. The mean values of the distributions are $\simeq 0\farcs1$ for both right ascension and declination.

We finally note that the internal accuracies of the catalogs used as reference in this section are very close to the measured dispersions, i.e.
$\sim0\farcs2$, $\sim0\farcs1$ and $\sim0\farcs07$ when comparing OmegaWINGS astrometry with to 2MASS, WINGS and SDSS, respectively.
We can therefore conclude that our astrometric calibration has an internal accuracy at the level of at least $0\farcs1$.

\subsection{Photometry}

\myfigure{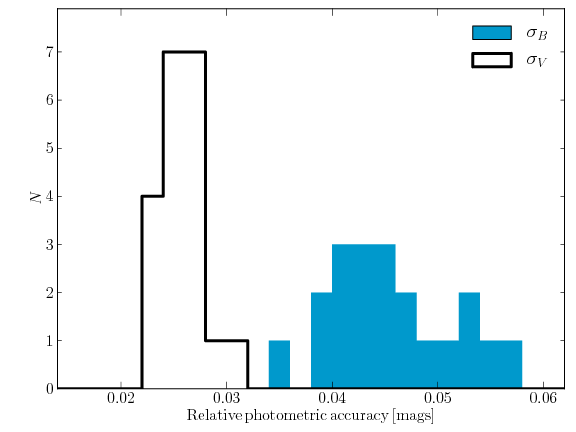}{Relative photometric accuracy based on a comparison with SDSS DR9.
The histograms shows the dispersion of the distributions of differences between OmegaWINGS and SDSS photometry.}{f:photsdss}

The relative accuracy of OmegaWINGS photometry
across the OmegaCAM FoV was tested by comparing OmegaWINGS
photometry with SDSS one.
We  adopted the linear colour equations proposed by 
\cite{jord+2006} to transform SDSS $ugr$  photometry into standard $BV$ magnitudes.
For each of the 20 OmegaWINGS field observed by SDSS,
we calculated the dispersion of the differences between OmegaWINGS and (transformed) SDSS magnitudes
for all stars with $B<20$ mag and $V<19$ mag. 
Results are shown in 
Figure \ref{f:photsdss}.
In the $V$-band the relative photometric accuracy is $\lesssim 0.03$ mag for all clusters.
The dispersions of $\Delta B$ are 0.04--0.06 mags. The systematically higher dispersion in the $B$- band
are due to non-linear colour terms in the  transformations from SDSS $ugr$ to $BV$ photometric systems
and/or a dependence of the transformations on stars metallicity/colour \citep{jord+2006}.
In Sect. \ref{sec:calib} in fact we found no high-order colour term in the comparison of OmegaWINGS photometry
with WINGS one. 
A detailed discussion of this issue is beyond the aim of this paper.

\begin{figure*}
  \centering
  \resizebox{.9\hsize}{!}{\includegraphics{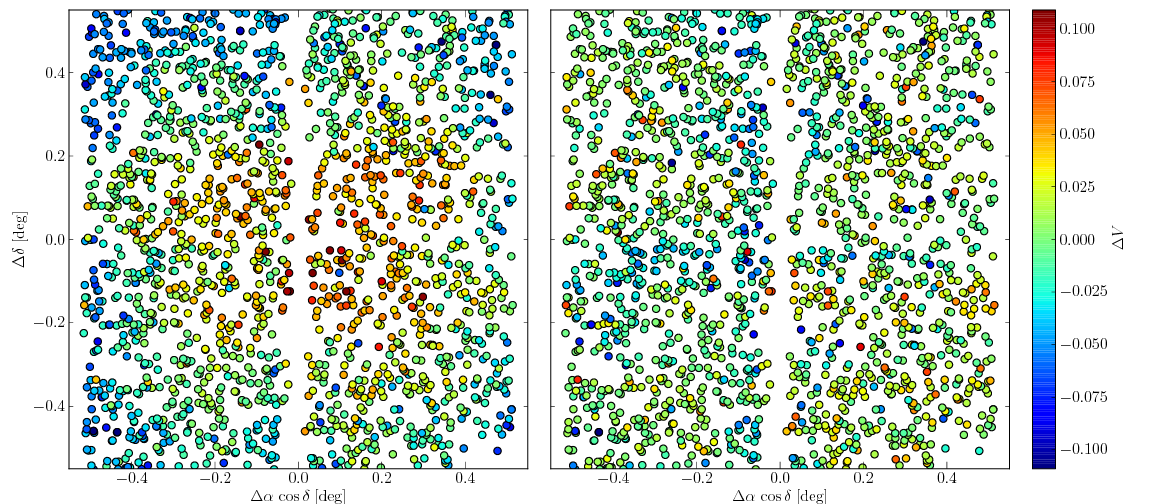}}
  \caption{
Magnitude difference between OmegaWINGS and SDSS photometry --transformed onto the Johnson's system--
for stars brighter than $V=20$ mag as a function of the position on the OmegaWINGS detector.
In the {\it right panel} the result is obtained from the final OmegaWINGS calibrated catalog, while
the photometry in the {\it left panel} was obtained with no illumination-correction.
}
  \label{f:illcor}
\end{figure*}

To check the spatial stability of OmegaWINGS calibration, we
analysed the magnitude difference between OmegaWINGS and (transformed) SDSS
photometry as a function of the position on the mosaic. 
As an example, the magnitude difference of all
stars with $V<19$ in Z8852 are shown in the right panel of Fig. \ref{f:illcor}.
For comparison, the same comparison when no illumination correction
is applied (see Sect. \ref{sec:datared})
 is shown in the left panel of Fig. \ref{f:illcor}.
 The maximum effect of the illumination correction, from the edge to the centre of the mosaic, is
$\sim 0.2$ mag.
We analysed  the same maps as the one shown in Fig. \ref{f:illcor} for all 20 OmegaWINGS fields
with available SDSS photometry in both 
$B$ and $V$-band and we conclude that
the photometric zero point in all calibrated catalogs is constant across the whole mosaic, 
and that there are no residual systematic effects of neither 
the illumination correction nor the gain harmonisation.

\begin{figure}
  \resizebox{\hsize}{!}{\includegraphics{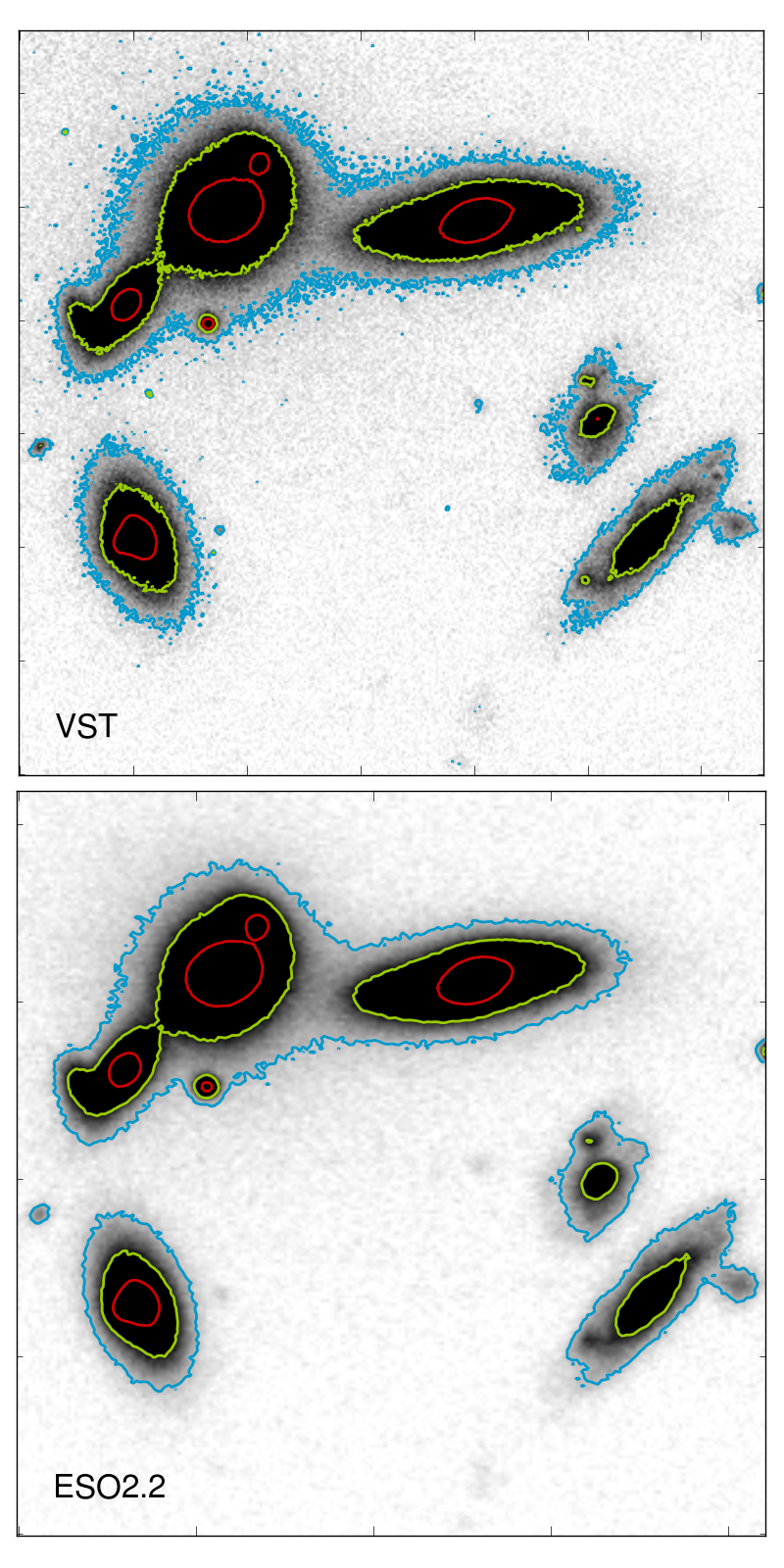}}
  \caption{
Zoom on a $35\arcsec\times35\arcsec$ region in the $V$-band VST image of A151 
and on the WINGS image, taken with the WFC@ESO2.2 telescope.
The same contour levels have been plotted in both images. Besides the
slightly lower S/N ratio of the
VST image with respect to ESO@2.2 image, the shape and location of the contour is the same.
This is an indication that the background subtraction did not alter the
faintest structures in galaxy haloes.}
\label{f:compsurf}
\end{figure}

\begin{figure}
  \resizebox{\hsize}{!}{\includegraphics{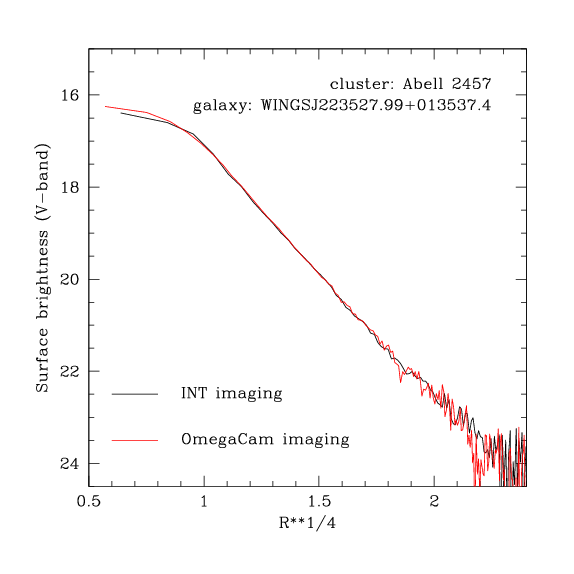}}
  \caption{
Comparison of the major-axis surface brightness profile of a bright galaxy
as obtained from OmegaCAM and WFC@INT images. The agreement of the two 
profiles demonstrates the reliability of our sky subtraction procedure.}
\label{f:trace}
\end{figure}

Finally, the sky-subtraction procedure has been tested by performing a detailed
analysis of the radial profiles of a few extended bright galaxies, and
comparing the results obtained from the final OmegaCAM stacked mosaics
with data from the WINGS survey.  As an example, in Fig \ref{f:compsurf} the
comparison of OmegaWINGS and WINGS images of a region populated by
several galaxies shows that the structure of the galactic haloes is the
same in the two images. This is a clear indication that the sky-subtraction
procedure did remove all large-scale artefacts from the images (see Fig. \ref{f:stack}),
but it did not alter the galaxy profiles.
A more quantitative analysis of this point can be derived from
the direct comparison of radial profiles of the same galaxy obtained
from OmegaWINGS and WINGS images, as the one presented in 
Fig. \ref{f:trace}. The profile obtained from OmegaCAM image is in
perfect agreement with the one obtained at the INT telescope, out to
the detection limit, corresponding to a radius of $\sim 16\arcsec$,
i.e. 75 pixels in OmegaCAM.
The minor differences in the central regions are due to the fact
that OmegaWINGS $V$-band observations of A2457 were carried out under better 
seeing conditions ($1\farcs1$)
than WINGS ones  ($1\farcs4$).

\subsection{Photometric completeness}\label{s:compl}

\begin{figure}
  \resizebox{\hsize}{!}{\includegraphics{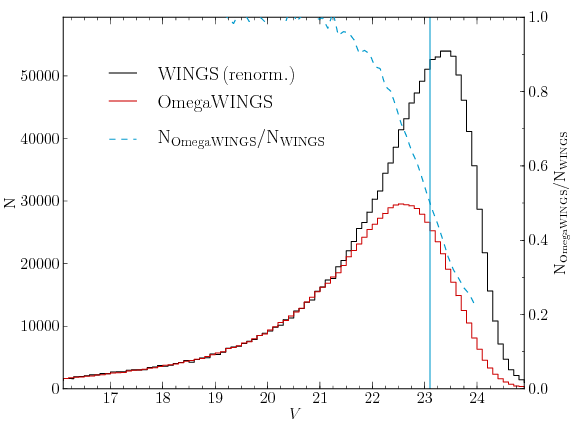}}
  \caption{
$V$-band magnitude distribution of all objects in the OmegaWINGS and WINGS database. 
WINGS MD was re-normalised to match the total number of sources with $16<V<21$ mag.
The ratio of OmegaWINGS to WINGS MDs is shown as a dotted line; the corresponding
scale is shown on the right-hand axis. The vertical is traced at the magnitude 
corresponding to a ratio of 0.5 (50\% completeness level).
}
\label{f:lfcompl}
\end{figure}

The overall OmegaWINGS photometric completeness factor was estimated
by comparing the magnitude distributions (MD) of all sources in OmegaWINGS and WINGS
catalogs. To perform the comparison, the WINGS distribution
was re-normalised 
to match the total number of OmegaWINGS sources with $16<V<21$ mag.
The MDs obtained for  all 45 OmegaWINGS fields are shown in Fig. \ref{f:lf_box}
in the appendix (available only on the online version of this paper).
The photometric depth depends on the seeing conditions during observations, but
OmegaWINGS photometry is in general 0.5--1.0 mag shallower than WINGS one.
However, when OmegaWINGS observations 
were carried out with seeing $\lesssim1\farcs0$, OmegaWINGS is as deep as (and in some cases deeper than) WINGS (see Fig.\ref{f:lf_box}).
The overall photometric depth of OmegaWINGS  was estimated by stacking together
all 45 MDs. Figure \ref{f:lfcompl} shows that OmegaWINGS MD peaks at
$V\sim 22.5$ mag and WINGS one at $V\sim 23.4$.
The OmegaWINGS completeness
can be estimated from as the ratio of OmegaWINGS to WINGS MDs.
The 50\% completeness level is reached at
$V=23.1$ mag, the 80\% level at 
$V=22.6$ mag (see Fig. \ref{f:lfcompl}).
This results is based on the assumption that WINGS photometry is complete at least up to $V\sim23$ mag.
We therefore performed a further test by
fitting an exponential relation to the bright tail of the histogram in 
Fig.  \ref{f:lfcompl}. The completeness factor was obtained as the ratio of the 
observed MD to the best-fit exponential model. Following this approach,
50\% and 80\% completeness are found at $V=23.3$ and 22.7 mag, respectively, confirming --within 
reasonable uncertainties-- our previous results.

\subsection{Star-galaxy classification}

\myfigure{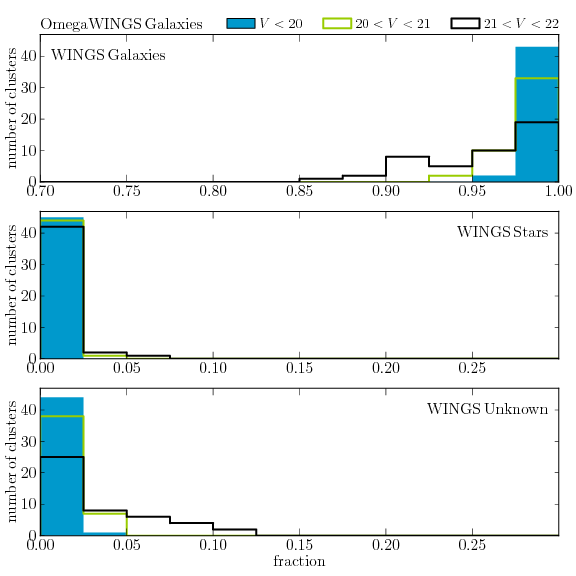}{Fraction of objects classified as galaxies in OmegaWINGS that have been 
classified as galaxies ({\it upper panel}), stars ({\it central panel}), or with unknown classification ({\it lower panel}) in WINGS.
In each panel solid blue,  green and black histograms show objects in three different magnitude bins, as indicated in the legend.
}{f:cl_omega}

To check the reliability of OmegaWINGS source classification, we compared it with the WINGS one.
For each cluster we divided the  sources classified as galaxies in OmegaWINGS in three magnitude bins: $V<20$ mag, $20<V<21$ mag, and $21<V<22$ mag.
We then calculated how many OmegaWINGS galaxies are classified as 
galaxy, star, and unknown  in WINGS. The histograms of these fractions are presented in 
Fig. \ref{f:cl_omega}.
 In all but two clusters the fraction of bright ($V<20$ mag) OmegaWINGS galaxies 
classified as galaxies in WINGS are $>97.5\%$.
In more than $50\%$ of the clusters the number of faint OmegaWINGS galaxies with unknown classification in WINGS
is negligible; in all other clusters, this fraction is still smaller than 10\% (see lower panel of Fig. \ref{f:cl_omega}).
In all clusters there are no OmegaWINGS galaxies with $V<20$ mag classified as stars in WINGS.
The fraction of galaxies misclassified as stars in WINGS is negligible ($<5\%$) also for fainter galaxies $20<V<22$).
To summarise, we can conclude that the classification of galaxies in OmegaWINGS is highly reliable.

The other question we addressed with this analysis is about the completeness of our classification, i.e. how many galaxies are missed
by our classification?
Clues on this issues can be provided by studying OmegaWINGS classification of WINGS galaxies, shown in 
Fig. \ref{f:cl_wings}.
In most clusters, all WINGS galaxies with $V<20$ mag are classified as galaxies also in OmegaWINGS,
just in a few clusters there are 
bright  WINGS galaxies otherwise classified in OmegaWINGS; their fraction is however always low ($5\%-10\%$).
The number of WINGS galaxies with $21<V<22$ mag with unknown OmegaWINGS classification
is not negligible in a significant number of clusters. 
Nonetheless, in $\sim 50\%$ of the clusters 
 the number of WINGS galaxies misclassified as stars in OmegaWINGS is $5-15\%$.
The reliability of the source classification is strongly dependent on seeing conditions during the observations, and in fact
the clusters with the most relevant discrepancies between OmegaWINGS and WINGS classification are those
where there are relevant seeing differences between the WINGS and OmegaWINGS observations.

To summarise, we conclude that OmegaWINGS source classification is highly reliable for all objects with $V<20$ mag. This is the magnitude
range used to select the targets for our AAOmega spectroscopic follow-up survey.
At faintest magnitudes, in clusters observed under non-optimal seeing conditions, the OmegaWINGS galaxy selection is not complete, i.e. a significant number of galaxies could likely be assigned an unknown classification. On the other hand, 
the classification of galaxies in OmegaWINGS is very robust, i.e. it is very unlikely that a OmegaWINGS galaxy is actually a star.

\myfigure{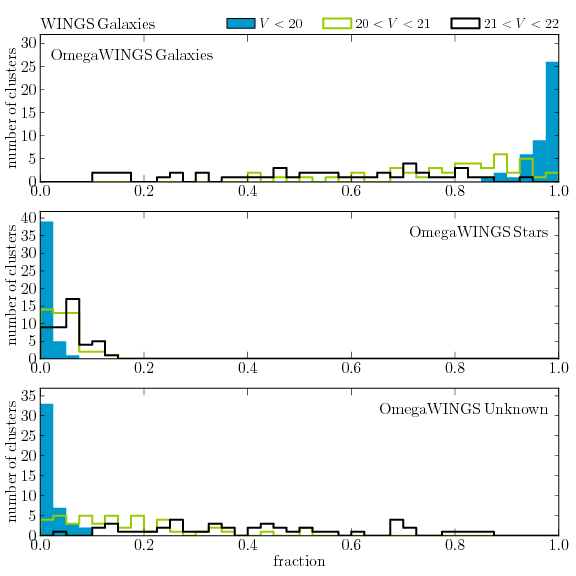}{
Fraction of objects classified as galaxies in WINGS that have been 
classified as galaxies ({\it upper panel}), stars ({\it central panel}), or with unknown classification ({\it lower panel}) in OmegaWINGS.
In each panel solid blue,  green and black histograms show objects in three different magnitude bins, as indicated in the legend.
}{f:cl_wings}

\section{Summary}\label{sec:summary}

This paper is the first of a series presenting OmegaWINGS, the
wider-field extension of the WINGS database for X-ray selected galaxy
clusters at $z=$0.04-0.07. The $B$- and $V$-band observations of the
46 WINGS clusters observed with OmegaCAM/VST are presented here, while
the ongoing $u$-band OmegaCAM/VST and spectroscopic AAOmega/AAT
follow-ups will be presented in subsequent papers.

All clusters have been observed for 25 min in each band, with a median
seeing of $1\arcsec$ in both $B$ and $V$-band, and $<1\farcs3$ and
$1\farcs2$ in 80\% of the $B$- and $V$-band images, respectively.  The
data have been reduced with a modified version of the ESO-MVM {\sc
  alambic} reduction package, developing ad hoc cross-talk, gain
harmonisation and CCD control procedures. Special care has been taken
for illumination correction, using purposely obtained OmegaCAM
observations of standard stellar fields.

Sextractor photometric catalogues have been produced and are released
with this paper at CDS. Catalogs and reduced images will also be part
of the next release version of the WINGS database.

The quality of the astrometry, photometric accuracy, star-galaxy
separation and sky-subtraction have been tested in various ways and
show that results are generally of the same or even better quality
than the previous WINGS results. The absolute astrometric accuracy is
$\sim 0\farcs2$ and $0\farcs07$ when the calibration is based on 2MASS
and SDSS, respectively. The photometric catalogues are 50\% complete 
at $V=23.1$ mag and 80\% complete at $V=22.6$ mag. 

The $B$- and $V$-band OmegaCAM images have provided the AAOmega
spectroscopic targets, and have been employed to identify jellyfish
candidate galaxies subject to ram pressure stripping (Poggianti et
al. submitted).  Ongoing analysis of these images include surface
brightness analysis with GASPHOT \citep{dono+2014} and morphological
classification with MORPHOT \citep{fasa+2012}. Together with the
spectroscopy, taking advantage of the large OmegaCAM field of view
and, consequently, of the large clustercentric radii reached, this
dataset is used to study the effects of the environment on galaxy
properties out to large distances from the cluster centre and for a
number of studies on the dynamical status and light distribution of
the clusters.

\begin{acknowledgements}
We would like to warmly thank Herv\'e Bouy and Benoit Vandame
for sharing their code and for the useful discussion.
We acknowledge financial support from the PRIN-INAF 2010 and 2014.
BV was supported by the World Premier International Research Center Initiative (WPI), MEXT, Japan and the Kakenhi Grant-in-Aid for Young Scientists (B)(26870140) from the Japan Society for the Promotion of Science (JSPS).
  This publication makes use of data products from the Two Micron All
  Sky Survey, which is a joint project of the University of
  Massachusetts and the Infrared Processing and Analysis
  Center/California Institute of Technology, funded by the National
  Aeronautics and Space Administration and the National Science
  Foundation.
  SDSS-III is managed by the Astrophysical Research Consortium for the Participating Institutions of the SDSS-III Collaboration including the University of Arizona, the Brazilian Participation Group, Brookhaven National Laboratory, Carnegie Mellon University, University of Florida, the French Participation Group, the German Participation Group, Harvard University, the Instituto de Astrofisica de Canarias, the Michigan State/Notre Dame/JINA Participation Group, Johns Hopkins University, Lawrence Berkeley National Laboratory, Max Planck Institute for Astrophysics, Max Planck Institute for Extraterrestrial Physics, New Mexico State University, New York University, Ohio State University, Pennsylvania State University, University of Portsmouth, Princeton University, the Spanish Participation Group, University of Tokyo, University of Utah, Vanderbilt University, University of Virginia, University of Washington, and Yale University.

\end{acknowledgements}
\bibliographystyle{aa} % style aa.bst
\bibliography{omegawings} % your references Yourfile.bib

\Online
\begin{appendix}
\section{Supplementary tables and figures}

\mytabletwo{l ll ll l}{
  cluster & 
  $a_B$ & $b_B$&
  $a_V$ & $b_V$&
  P}{
  A119       & $  0.116 \pm   0.004 $ & $ 24.291 \pm   0.006$ & $  0.030 \pm   0.004 $ & $ 24.291 \pm   0.006$ & P88 \\ 
A160       & $  0.116 \pm   0.004 $ & $ 24.405 \pm   0.005$ & $  0.030 \pm   0.004 $ & $ 24.295 \pm   0.005$ & P88 \\ 
A193       & $  0.116 \pm   0.004 $ & $ 24.367 \pm   0.005$ & $  0.030 \pm   0.004 $ & $ 24.313 \pm   0.005$ & P88 \\ 
A3128      & $  0.116 \pm   0.004 $ & $ 24.221 \pm   0.004$ & $  0.030 \pm   0.004 $ & $ 24.266 \pm   0.004$ & P88 \\ 
A3158      & $  0.116 \pm   0.004 $ & $ 24.312 \pm   0.004$ & $  0.030 \pm   0.004 $ & $ 24.213 \pm   0.004$ & P88 \\ 
A500       & $  0.116 \pm   0.004 $ & $ 24.145 \pm   0.004$ & $  0.030 \pm   0.004 $ & $ 24.226 \pm   0.004$ & P88 \\ 
A754       & $  0.116 \pm   0.004 $ & $ 24.320 \pm   0.003$ & $  0.030 \pm   0.004 $ & $ 24.340 \pm   0.004$ & P88 \\ 
A957x      & $  0.116 \pm   0.004 $ & $ 24.313 \pm   0.005$ & $  0.030 \pm   0.004 $ & $ 24.311 \pm   0.005$ & P88 \\ 
A970       & $  0.116 \pm   0.004 $ & $ 24.281 \pm   0.004$ & $  0.030 \pm   0.004 $ & $ 24.333 \pm   0.004$ & P88 \\ 
A1983      & $  0.128 \pm   0.003 $ & $ 24.297 \pm   0.004$ & $  0.040 \pm   0.003 $ & $ 24.250 \pm   0.004$ & P89 \\ 
A2382      & $  0.128 \pm   0.003 $ & $ 24.294 \pm   0.003$ & $  0.040 \pm   0.003 $ & $ 24.293 \pm   0.004$ & P89 \\ 
A2399      & $  0.128 \pm   0.003 $ & $ 24.357 \pm   0.003$ & $  0.040 \pm   0.003 $ & $ 24.267 \pm   0.003$ & P89 \\ 
A2415      & $  0.128 \pm   0.003 $ & $ 24.310 \pm   0.003$ & $  0.040 \pm   0.003 $ & $ 24.321 \pm   0.004$ & P89 \\ 
A2457      & $  0.128 \pm   0.003 $ & $ 24.316 \pm   0.004$ & $  0.040 \pm   0.003 $ & $ 24.287 \pm   0.004$ & P89 \\ 
A3556      & $  0.128 \pm   0.003 $ & $ 24.227 \pm   0.003$ & $  0.040 \pm   0.003 $ & $ 24.235 \pm   0.003$ & P89 \\ 
A3560      & $  0.128 \pm   0.003 $ & $ 24.299 \pm   0.003$ & $  0.040 \pm   0.003 $ & $ 24.217 \pm   0.003$ & P89 \\ 
A3809      & $  0.128 \pm   0.003 $ & $ 24.259 \pm   0.003$ & $  0.040 \pm   0.003 $ & $ 24.204 \pm   0.003$ & P89 \\ 
MKW3s      & $  0.128 \pm   0.003 $ & $ 24.255 \pm   0.003$ & $  0.040 \pm   0.003 $ & $ 24.231 \pm   0.003$ & P89 \\ 
A151       & $  0.122 \pm   0.003 $ & $ 24.316 \pm   0.005$ & $  0.036 \pm   0.004 $ & $ 24.229 \pm   0.005$ & P90 \\ 
A1631a     & $  0.122 \pm   0.003 $ & $ 24.377 \pm   0.004$ & $  0.036 \pm   0.004 $ & $ 24.285 \pm   0.004$ & P90 \\ 
A2593      & $  0.122 \pm   0.003 $ & $ 24.275 \pm   0.004$ & $  0.036 \pm   0.004 $ & $ 24.342 \pm   0.004$ & P90 \\ 
A3266      & $  0.122 \pm   0.003 $ & $ 24.259 \pm   0.003$ & $  0.036 \pm   0.004 $ & $ 24.249 \pm   0.004$ & P90 \\ 
A3395      & $  0.122 \pm   0.003 $ & $ 23.825 \pm   0.003$ & $  0.036 \pm   0.004 $ & $ 23.967 \pm   0.003$ & P90 \\ 
A3376      & $  0.122 \pm   0.003 $ & $ 24.233 \pm   0.003$ & $  0.036 \pm   0.004 $ & $ 24.164 \pm   0.003$ & P90 \\ 
Z8852      & $  0.122 \pm   0.003 $ & $ 24.310 \pm   0.004$ & $  0.036 \pm   0.004 $ & $ 24.261 \pm   0.004$ & P90 \\ 
A1069      & $  0.133 \pm   0.002 $ & $ 24.321 \pm   0.003$ & $  0.044 \pm   0.002 $ & $ 24.336 \pm   0.003$ & P91 \\ 
A147       & $  0.133 \pm   0.002 $ & $ 24.338 \pm   0.004$ & $  0.044 \pm   0.002 $ & $ 24.311 \pm   0.004$ & P91 \\ 
A168       & $  0.133 \pm   0.002 $ & $ 24.340 \pm   0.005$ & $  0.044 \pm   0.002 $ & $ 24.299 \pm   0.005$ & P91 \\ 
A1991      & $  0.133 \pm   0.002 $ & $ 24.304 \pm   0.004$ & $  0.044 \pm   0.002 $ & $ 24.283 \pm   0.004$ & P91 \\ 
A2107      & $  0.133 \pm   0.002 $ & $ 24.407 \pm   0.003$ & $  0.044 \pm   0.002 $ & $ 24.356 \pm   0.003$ & P91 \\ 
A2589      & $  0.133 \pm   0.002 $ & $ 24.326 \pm   0.004$ & $  0.044 \pm   0.002 $ & $ 24.324 \pm   0.004$ & P91 \\ 
A2657      & $  0.133 \pm   0.002 $ & $ 24.286 \pm   0.004$ & $  0.044 \pm   0.002 $ & $ 24.317 \pm   0.004$ & P91 \\ 
A2665      & $  0.133 \pm   0.002 $ & $ 24.333 \pm   0.004$ & $  0.044 \pm   0.002 $ & $ 24.295 \pm   0.004$ & P91 \\ 
A2717      & $  0.133 \pm   0.002 $ & $ 24.304 \pm   0.005$ & $  0.044 \pm   0.002 $ & $ 23.847 \pm   0.005$ & P91 \\ 
A2734      & $  0.133 \pm   0.002 $ & $ 24.343 \pm   0.004$ & $  0.044 \pm   0.002 $ & $ 24.296 \pm   0.004$ & P91 \\ 
A3528      & $  0.133 \pm   0.002 $ & $ 24.277 \pm   0.002$ & $  0.044 \pm   0.002 $ & $ 24.196 \pm   0.002$ & P91 \\ 
A3530      & $  0.133 \pm   0.002 $ & $ 24.298 \pm   0.002$ & $  0.044 \pm   0.002 $ & $ 24.231 \pm   0.003$ & P91 \\ 
A3532      & $  0.133 \pm   0.002 $ & $ 24.292 \pm   0.002$ & $  0.044 \pm   0.002 $ & $ 24.212 \pm   0.003$ & P91 \\ 
A3558      & $  0.133 \pm   0.002 $ & $ 24.253 \pm   0.002$ & $  0.044 \pm   0.002 $ & $ 24.223 \pm   0.002$ & P91 \\ 
A3667      & $  0.133 \pm   0.002 $ & $ 24.247 \pm   0.002$ & $  0.044 \pm   0.002 $ & $ 24.257 \pm   0.002$ & P91 \\ 
A3716      & $  0.133 \pm   0.002 $ & $ 24.311 \pm   0.002$ & $  0.044 \pm   0.002 $ & $ 24.265 \pm   0.003$ & P91 \\ 
A3880      & $  0.133 \pm   0.002 $ & $ 23.841 \pm   0.004$ & $  0.044 \pm   0.002 $ & $ 24.289 \pm   0.004$ & P91 \\ 
A4059      & $  0.133 \pm   0.002 $ & $ 24.286 \pm   0.004$ & $  0.044 \pm   0.002 $ & $ 24.250 \pm   0.004$ & P91 \\ 
A85        & $  0.133 \pm   0.002 $ & $ 24.372 \pm   0.004$ & $  0.044 \pm   0.002 $ & $ 24.315 \pm   0.004$ & P91 \\ 
IIZW108    & $  0.133 \pm   0.002 $ & $ 24.325 \pm   0.002$ & $  0.044 \pm   0.002 $ & $ 24.278 \pm   0.002$ & P91 \\ 

}{Photometric zero points and colour terms.}{tab:calib}

\mytabletwo{lcl}{
column&units&description}{
WINGS-ID&&unique identificator\\
cluster name&&name of host cluster\\
RA(J2000)    &deg& Right Ascension of emission peak\\
DEC(J2000) &deg& Declination of emission peak\\
Area& arcsec$^2$&Isophotal area\\
KronRad &arcsec&Kron radius\\
FWHM&arcsec&Full width at half maximum along major axis\\
$b/a$&&axial ratio\\
PA&deg&[-90/90] Position angle (North=0, Eastwards)\\
$\mu_B^0$&mag arcsec$^{-2}$&$B$-band surface brightness of the emission peak\\
$\mu_V^0$&mag arcsec$^{-2}$&$V$-band surface brightness of the emission peak\\
$B_\textrm{ISO}$   &mag  & $B$-band SExtractor's MAG\_ISO\\
$V_\textrm{ISO}$   &mag  & $V$-band SExtractor's MAG\_ISO\\
$B_\textrm{ISOC}$  &mag  & $B$-band SExtractor's MAG\_ISOC\\
$V_\textrm{ISOC}$  &mag  & $V$-band SExtractor's MAG\_ISOC\\
$B_\textrm{AUTO}$  &mag  & $B$-band SExtractor's MAG\_AUTO\\
$V_\textrm{AUTO}$  &mag  & $V$-band SExtractor's MAG\_AUTO\\
$B_\textrm{2pix}$  &mag  & $B$-band magnitude in aperture of 2 pixels\\
$V_\textrm{2pix}$  &mag  & $V$-band magnitude in aperture of 2 pixels\\
$B_\textrm{5pix}$  &mag  & $B$-band magnitude in aperture of 5 pixels\\
$V_\textrm{5pix}$  &mag  & $V$-band magnitude in aperture of 5 pixels\\
$B_\textrm{10pix}$ &mag  & $B$-band magnitude in aperture of 10 pixels\\
$V_\textrm{10pix}$ &mag  & $V$-band magnitude in aperture of 10 pixels\\
$B_\textrm{5pix}$  &mag  & $B$-band magnitude in aperture of 5 pixels\\
$V_\textrm{5pix}$  &mag  & $V$-band magnitude in aperture of 5 pixels\\
$B_\textrm{10pix}$ &mag  & $B$-band magnitude in aperture of 10 pixels\\
$V_\textrm{10pix}$ &mag  & $V$-band magnitude in aperture of 10 pixels\\
$B_\textrm{4kpc}$  &mag  & $B$-band magnitude in aperture of 4 kpc\\
$V_\textrm{4kpc}$  &mag  & $V$-band magnitude in aperture of 4 kpc\\
$B_\textrm{10kpc}$  &mag  & $B$-band magnitude in aperture of 10 kpc\\
$V_\textrm{10kpc}$  &mag  & $V$-band magnitude in aperture of 10 kpc\\
$B_\textrm{20kpc}$ &mag  & $B$-band magnitude in aperture of 20 kpc\\
$V_\textrm{20kpc}$ &mag  & $V$-band magnitude in aperture of 20 kpc\\
$B_\textrm{fib1}$  &mag  & $B$-band magnitude in aperture of $1\farcs60$\\
$V_\textrm{fib1}$  &mag  & $V$-band magnitude in aperture of $1\farcs60$\\
$B_\textrm{fib2}$  &mag  & $B$-band magnitude in aperture of $2\farcs00$\\
$V_\textrm{fib2}$  &mag  & $V$-band magnitude in aperture of $2\farcs00$\\
$B_\textrm{fib3}$  &mag  & $B$-band magnitude in aperture of $2\farcs16$\\
$V_\textrm{fib3}$  &mag  & $V$-band magnitude in aperture of $2\farcs16$\\
}{Description of the table available at CDS.
}{tab:tabdesc}

\begin{figure*}
  \centering
  \resizebox{.95\hsize}{!}{\includegraphics{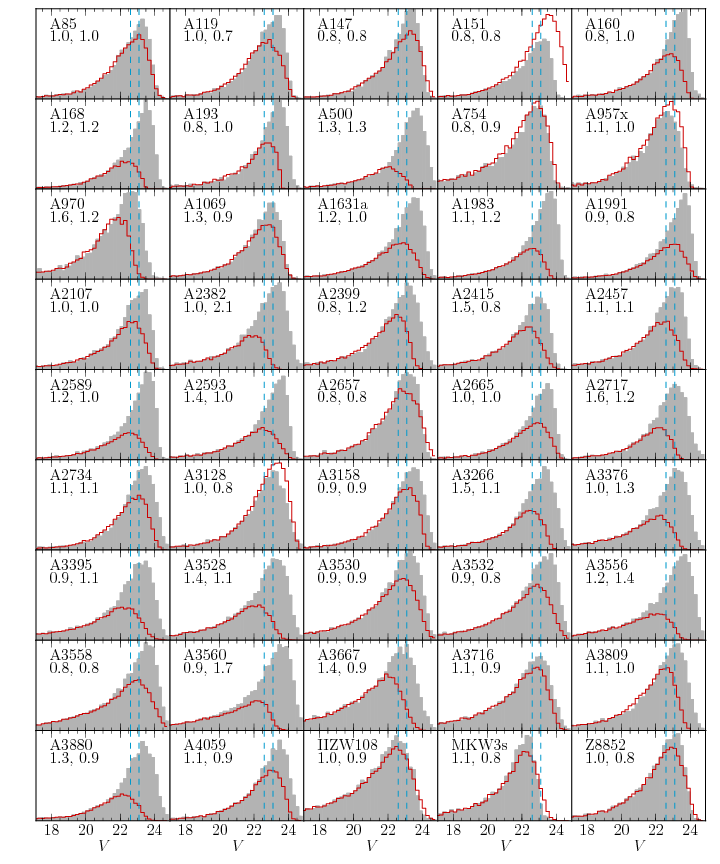}}
  \caption{
Magnitude distributions of all sources in OmegaWINGS ({\it red histograms})
and WINGS ({\it shaded grey histograms}) for all 45 OmegaWINGS fields.
WINGS MDs are renormalised to match  the number of bright ($16<V<21$ mag) objects  in OmegaWINGS LFs.
The vertical lines show the overall OmegaWINGS 50\% and 80\% completeness level (see Sect. \ref{s:compl}).
The label in each panel indicates the seeing in $B$- and $V$-band images, in arcseconds.}
  \label{f:lf_box}
\end{figure*}

\end{appendix}
\end{document}